# Artificial Intelligence-Enabled Holistic Design of Catalysts Tailored for Semiconducting Carbon Nanotube Growth

Liu Qian,[1, 2 §*] Yue Li,[1 §] Ying Xie,[1, 2] Jian Zhang,[3] Pai Li,[4] Yue Yu,[2] Zhe Liu,[2] Feng Ding,[5] Jin Zhang[1, 2 *]

[1]School of Materials Science and Engineering, Peking University, Beijing 100871, China.

[2]Beijing Science and Engineering Center for Nanocarbons, Beijing National Laboratory for Molecular Sciences, College of Chemistry and Molecular Engineering, Peking University, Beijing 100871, China.

[3]Nanofabrication Laboratory, National Center for Nanoscience and Technology, Beijing 100190, China

[4]State Key Laboratory of Materials for Integrated Circuits, Shanghai Institute of Microsystem and Information Technology, Chinese Academy of Sciences, Shanghai 200050, China

[5] Suzhou Laboratory, Suzhou, Jiangsu 215123, China

[§] These authors contributed equally to this work: Liu Qian and Yue Li

* E-mail: jinzhang@pku.edu.cn; qianliu-cnc@pku.edu.cn

**Abstract**

Catalyst design is crucial for materials synthesis, especially for complex reaction networks. Strategies like collaborative catalytic systems and multifunctional catalysts are effective but face challenges at the nanoscale. Carbon nanotube synthesis contains complicated nanoscale catalytic reactions, thus achieving high-density, high-quality semiconducting CNTs demands innovative catalyst design. In this work, we present a holistic framework integrating machine learning into traditional catalyst design for semiconducting CNT synthesis. It combines knowledge-based insights with data-driven techniques. Three key components, including open-access electronic structure databases for precise physicochemical descriptors, pre-trained natural language processing-based embedding model for higher-level abstractions, and physical - driven predictive models based on experiment data, are utilized. Through this framework, a new method for selective semiconducting CNT synthesis via catalyst - mediated electron injection, tuned by light during growth, is proposed. 54 candidate catalysts are screened, and three with high potential are identified. High-throughput experiments validate the predictions, with semiconducting selectivity exceeding 91% and the $FeTiO_3$ catalyst reaching 98.6%. This approach not only addresses semiconducting CNT synthesis but also offers a generalizable methodology for global catalyst design and nanomaterials synthesis, advancing materials science in precise control.

**Introduction**

Catalyst design plays a pivotal role in advancing materials synthesis, where precise control over chemical reactions is crucial for achieving desired structural and functional properties[1]. Addressing the complexity of chemical reaction networks often involves two main strategies: employing collaborative catalytic systems, where multiple catalysts work in tandem to drive distinct reaction pathways, and designing multifunctional catalysts capable of simultaneously performing diverse roles within a single material. These approaches have proven effective in enabling complex reaction networks at larger scales, providing tools to steer reaction selectivity, improve efficiency, and tailor product properties[2]. However, translating these strategies to the nanoscale introduces unique challenges, as the confined reaction environments demand even greater precision and integration of catalytic functionality[3].

Carbon nanotube (CNT) synthesis exemplifies such a complex system. CNTs are among the most promising materials for next-generation electronic devices due to their exceptional electrical, thermal, and mechanical properties[4-6]. However, the realization of these applications hinges on the ability to produce semiconducting CNTs (s-CNTs) with high density, high quality, and good alignment[7]. Achieving this level of precision places extraordinary demands on the design of catalysts for horizontally aligned CNT (HACNT) array synthesis. Catalysts for semiconducting HACNT arrays must simultaneously fulfill multiple roles: decomposing the carbon source, serving as a template for tube structure, and facilitating selective growth of semiconducting CNTs-often by providing etchants or inducing specific structural epitaxy[8,9]. Compounding these challenges is the nanoscale nature of the catalysts themselves, which are typically several nanometers in size. At this scale, the catalysts must not only meet the functional requirements but also be manufacturable and stable under harsh growth conditions. These intertwined demands make designing such multifunctional catalysts a formidable task, requiring innovative approaches that extend beyond conventional design paradigms.

Recent advancements in applying artificial intelligence (AI) to scientific discovery, especially through machine learning (ML) techniques, have demonstrated their potential to analyze and optimize complex scientific systems[10], such as atmospheric systems[11], nuclear fusion control[12], materials synthesis[13,14], and optimization[15,16]. These approaches offer the ability to uncover hidden patterns and relationships within vast, multidimensional datasets, enabling more informed decisions in catalyst design[17]. However, applying AI methods to the intricately linked processes of nanomaterial synthesis remains underexplored, especially in systems demanding simultaneous optimization of multiple physical and chemical phenomena.

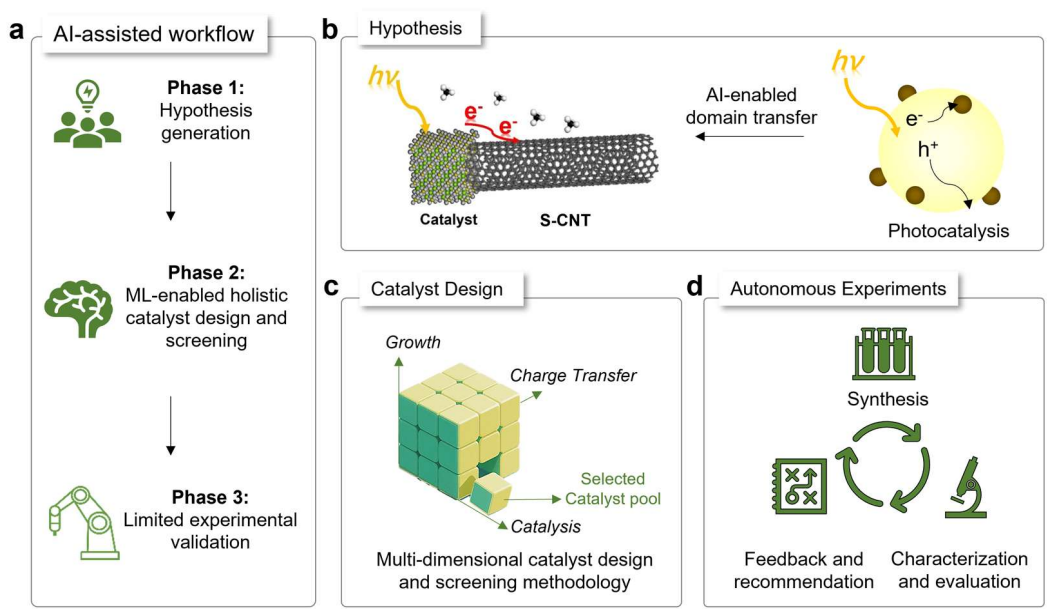

**Fig. 1 Holistic framework for catalyst design. a**, Schematic of the AI-assisted workflow. **b**, Schematic of the hypothesis: electron transfer mediated by photo-induced electrons enhances semiconducting CNT growth. **c**, The architecture for holistic, multi-dimensional evaluation of catalysts, including the catalysis ability, the electron transfer ability and the CNT growth ability. **d**, Schematic of the high-throughput autonomous experiments, which contains a self- optimization process with active learning.

In this work, we developed a holistic framework that integrates AI methods into traditional catalyst design methodologies to address the challenges of semiconducting CNT synthesis (**Figure 1a**). Our approach combines knowledge-based insights with data-driven techniques. Initially, we utilized simulations and domain expertise to generate scientific hypotheses of photo-induced electron transfer from catalyst to CNT (Fig. 1b), and defined the screening space for catalyst screening. We systematically evaluate the catalyst within a multi-dimensional space using integrated ML models (Fig. 1c). Notably, domain knowledge from photocatalysis to the catalyst design for CNT growth is successfully transferred, demonstrating the advantages of AI methods in knowledge transfer and generalization (Fig. 1b). This integrated design paradigm enables rational performance ranking of candidate catalysts, where subsequent minimal experimental verification, containing active learning strategy, efficiently identifies optimal catalyst-material parameter configurations, significantly improving the research efficiency compared to traditional trial-and-error methods (Fig. 1d).

In the ML models, we incorporated three key components: open-access electronic structure databases (from literatures) to provide specific descriptors of precise physicochemical processes, pre-trained natural language processing (NLP)-based embedding models (mat2vec)[18,19] to generate higher-level abstractions capturing generalized physicochemical processes, and physical-driven predictive models trained on a dataset of over 700 high-quality experimental results to bridge scientific hypotheses with experimental outcomes, ensuring experimental feasibility is considered at the design stage. Using this framework, we efficiently screened 54 candidate catalysts, identifying three with exceptional potential for achieving high-purity semiconducting CNT arrays. These predictions were validated using automated experimental platforms. All three selected catalysts, after minimal optimization, successfully produced high-density HACNT arrays with over 91% semiconducting selectivity. Notably, one catalyst achieved HACNT arrays with a semiconducting selectivity of 98.6%, exemplifying the framework's ability to optimize catalysts for complex systems. Beyond semiconducting CNT synthesis, our approach offers a generalizable methodology for addressing multifaceted challenges in nanomaterials synthesis.

## Results and Discussion

### Design and Screening of Catalysts

The fundamental difference between metallic and semiconducting CNTs lies in their electronic structure, a distinction that is inherently manifested in electron transport behavior. Consequently, controlling electron transfer during their synthesis offers a powerful strategy for achieving selective growth. Previous reports have highlighted charging behaviors during CNT growth[20,21], which are potentially linked to charge transfer interactions between the carbon source, catalyst, and CNTs. The density functional theory (DFT) simulations (For details see Methods) verified that electron accumulation on CNTs can significantly reduce the formation energy for CNT growth[22] (Supplementary Fig. 1). The high conductivity of metallic CNTs hinders the localization of injected electrons at the growth site, thereby amplifying the energetic disparity between semiconducting and metallic CNTs. We thus hypothesized that promoting electron transfer between the catalysts and CNTs could preferentially enhance the growth of semiconducting CNTs.

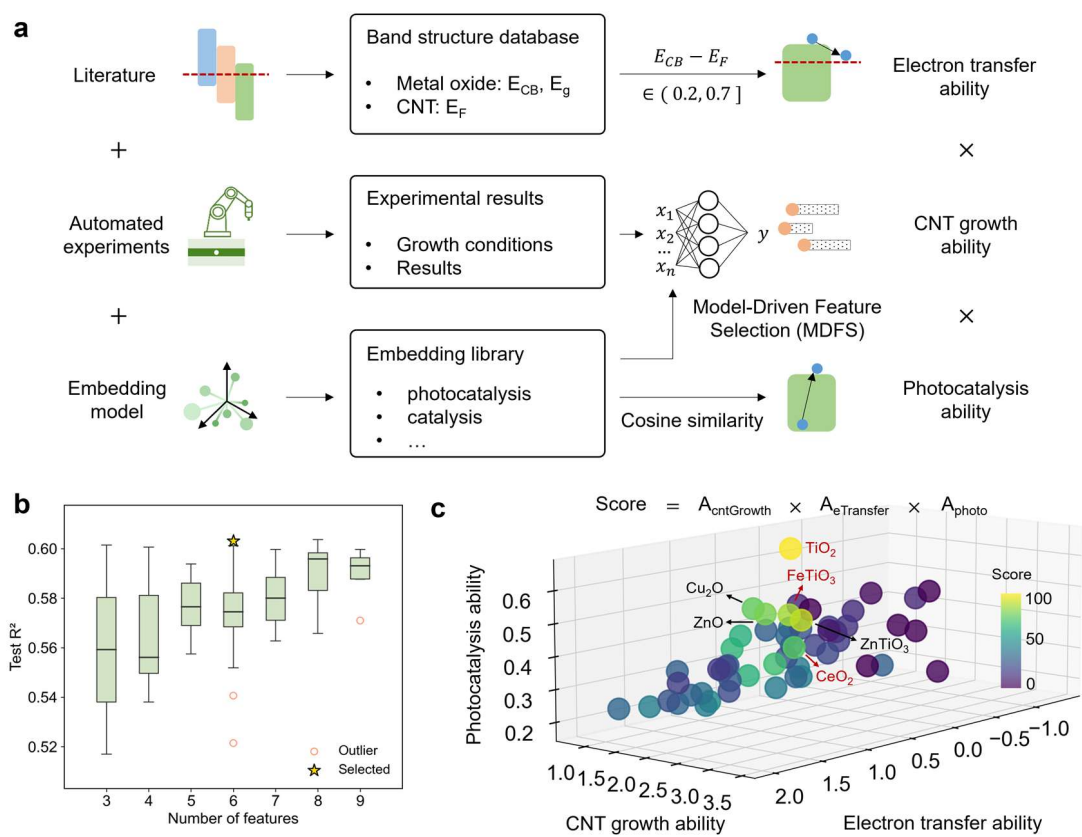

**Fig. 2 Methods and results of the holistic catalyst design. a**, Detailed methods for holistic evaluation of catalysts. **b**, Model performance (Test $R^2$) as a function of the number of descriptors used. The optimal set of six descriptors (marked by a star) achieved the best predictive accuracy with balanced simplicity. **c**, Scatter plot of the evaluation results based on their electron transfer, CNT growth, and photocatalytic abilities. Top-performing catalysts are highlighted by their scores.

To produce charge transfer during synthesis, introducing an external field can be an effective way. Given the practical challenges of applying external fields during high-temperature growth processes, we opted for a light field. This approach necessitates the design of a catalyst capable of generating photo-induced carriers, transferring electrons efficiently to CNTs, and simultaneously supporting their growth. Based on this, we developed a catalysts screening framework, schematically depicted in **Figure 2a**. Among the three critical requirements for catalyst design, electron transfer is the most straightforward to evaluate. For effective exiton dissociation and electron transfer, the conduction band edge of catalysts ($E_{CB}$) must be higher than the CNT Fermi level ($E_F$)[23] within the range of (0.2, 0.7] eV[24]. This range ensures sufficient energy alignment for electron transfer while minimizing the required driving force. Using open-access band structure databases, we screened metal oxides based on $E_{CB}$ and bandgap ($E_g$), identifying catalysts with optimal electron transfer properties.

In contrast, evaluating photocatalytic potential, which involves light absorption, charge separation, and charge recombination, is far more complex. These coupled physicochemical processes lack a direct descriptor for comprehensive evaluation. To address this, we adopted a knowledge-driven approach, leveraging the mat2vec embedding model. This model, trained on a vast corpus of materials science abstracts, encodes scientific phrases as high-dimensional vectors. By calculating the cosine similarity between the term "photocatalysis" and candidate catalysts, we derived a generalized descriptor representing photocatalytic potential. This approach effectively incorporates knowledge from the broader photocatalysis field, enabling systematic evaluation of candidate materials (For details see Supplementary Note 1).

The final and most critical criterion is the ability of catalysts to promote CNT growth. This process encompasses a series of intricate physicochemical interactions, including carbon source decomposition, structural templating, and alignment control. To tackle these challenges, we developed a combined knowledge- and data-driven framework to evaluate the feasibility of CNT growth under varying conditions. For the data part, we utilized the high-quality growth outcome database from our previously developed automated platform, Carbon Copilot (CARCO)[25]; however, the key challenge is extrapolating from this data to predict outcomes for unexplored catalyst candidates. Word embedding techniques can bridge knowledge across different scientific domains; therefore, we employed mat2vec and calculated cosine similarities between 12 growth-related descriptors and catalysts in the database. These descriptors were designed to replace nominal catalyst compositions in the dataset. To identify the most effective descriptors, we randomly evaluated 100 combinations of these 12 descriptors by training multi-layer perceptron (MLP) models on the modified dataset. The $R^2$ score of these models served as the metric for descriptor effectiveness.

Fig. 2b illustrates the model performance (Test $R^2$) as a function of the number of descriptors used. A specific combination of six descriptors (marked by a star) was ultimately selected, as it provided highest predictive power while maintaining model simplicity (Supplementary Fig. 6). With these descriptors, we calculated cosine similarities for all candidate catalysts and conducted virtual experiments using the trained MLP model. By averaging predicted growth outcomes across sampled parameter spaces, we derived an index to represent the feasibility of achieving CNT growth (For details see Supplementary Note 2). This index facilitated the ranking of candidate catalysts, emphasizing those with the greatest potential for high-quality CNT array synthesis.

Based on the design methods above, we finally defined a scoring mechanism combining three catalyst abilities. The results in Fig. 2c present the ranking of the photocatalysts, which were designed by comprehensively considering the electron transfer ability, photocatalysis ability, and CNT growth ability. From 54 screened candidates, 6 catalysts were selected for further validation.

**High-throughput validation experiments**

Based on the CARCO platform, we conducted a series of validation experiments. According to our design, we need to introduce light into the CVD system during growth. Benefiting from the quartz observation window of the furnace equipment in the CARCO platform, we could introduce xenon lamp light directly into the tube through the window. To introduce the light source more effectively, a lens was added ahead of the quartz window for convergence of the light source, which ultimately ensured that the light source hit the growth substrate with a suitable spot size. The xenon lamp was mainly used to cover a wide spectral range of light sources from the UV to the visible, which can cover the bandgaps of most optical semiconducting materials.

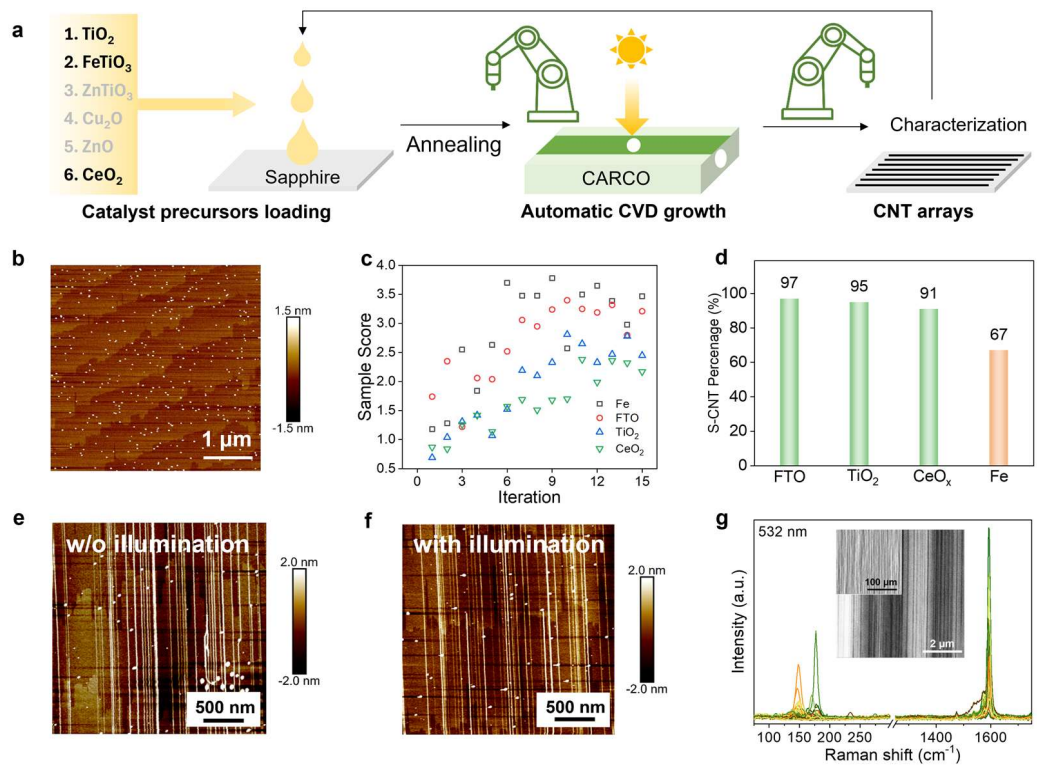

**Fig. 3 Schematic and results of the high-throughput validation experiments.**

**a**, Schematic demonstrating the whole process of validation experiments, including the loading of catalyst precursors, the automatic growth of CNT arrays and the characterization and optimization of CNT arrays. **b**, A typical AFM image of catalyst nanoparticles dispersed on the sapphire substrate. **c**, Sample scores of different catalysts over autonomous optimization iterations. **d**, A histogram of the percentage of semiconducting CNTs grown from different catalysts with illumination (statistics from Raman spectra). **e-f**, Typical AFM images of HACNT arrays grown from FTO catalyst without (**e**) and with (**f**) illumination. **g**, Raman spectra of HACNT arrays grown from FTO catalyst with illumination. Inset is the corresponding high-magnification SEM image.

Among the top six catalysts predicted by our AI-enabled workflow, $TiO_2$, $FeTiO_3$ (FTO), and $CeO_2$ were selected, considering the synthesis feasibility and stability under high-temperature and reductive atmosphere. The Fe catalyst was used as the control group. A typical process of our validation experiments is schematically shown in **Figure 3a**. To obtain high-purity horizontally aligned SWCNT arrays on the substrate, catalyst nanoparticles were first prepared on sapphire substrates via the sol-gel method and calcination. $Fe(OH)_3$, $TiO_2$, FTO, and $Ce(NO)_3$ solutions were first synthesized as the precursors of Fe, $TiO_2$, FTO, and $CeO_2$ catalysts, respectively. We spin-coated the precursor onto the sapphire substrates, followed by high-temperature (up to 1100°C) calcination in the air atmosphere. Then, the substrates with catalyst-nanoparticle loading were ready for automatic CVD growth (For details see Methods).

X-ray photoelectron spectroscopy (XPS) was used to analyze the chemical composites of the catalysts on the substrate after calcination. The Ti 2p spectra (Supplementary Fig. 7a) containing two peaks are characteristic of Ti in the +4 oxidation state, confirming the formation of $TiO_2$ after spin-coating and calcination for $TiO_2$-samples. The Fe 2p spectra in Supplementary Fig. 7b show the presence of $Fe^{3+}$ in addition to $Fe^{2+}$ in the FTO-samples. Combined with the Raman spectrum in Supplementary Fig. 8, it is concluded that the products on the substrate are mainly FTO, and $Fe_2O_3$ may also be present, which is related to the unavoidable oxidization process during the high-temperature process[26]. In the next step of CNT growth, hydrogen can reduce the iron oxides, and these partially oxidized components from FTO can be reduced. For $CeO_2$-samples, the Ce 3d spectra exhibited multiple peaks due to the presence of multiple oxidation states[27] (Supplementary Fig. 7c). Cerium oxides are easy to be reduced under $H_2$ atmosphere during growth, thus the Cerium oxide catalysts are marked as $CeO_x$ in the following. The concentration of the catalyst precursor solution was optimized to obtain catalyst nanoparticles suitable for CNT growth after calcination and $H_2$ reduction. A typical morphology of the FTO nanoparticles is shown in the atomic force microscopy (AFM) image in Fig. 3b, which demonstrates nanoparticles uniformly distributed on the substrate. More AFM images and size statistics results are shown in Supplementary Fig. 9.

Fig. 3c illustrates the Bayesian optimization process based on the sample density implemented on the CARCO platform, where each iteration updates growth parameters based on feedback from characterization results. This closed-loop strategy gradually improves sample scores across catalyst iterations, demonstrating the effectiveness of automated optimization. Under these conditions, HACNT arrays with high density and good alignment were obtained, as confirmed by SEM images in Supplementary Fig. 10. Using the specific optimum for each catalyst, we then introduced illumination during growth. After careful Raman line mapping under multiple excitation wavelengths, the percentage of semiconducting CNTs relative to the Fe catalyst was quantified by statistical analysis of resonant RBM peaks, with the results summarized in Fig. 3d (see original Raman spectra of RBM region in Supplementary Figs. 11-15). All three catalysts predicted by our holistic framework showed great ability in selective growth of s-CNTs under illumination, all higher than 91%, which aligned well with the AI-enabled design. The highest purity of 97% was achieved by FTO-samples compared with a purity of 67% achieved by the Fe catalyst. The detailed optimization process and growth mechanism are further elucidated below, using the FTO catalyst as a representative.

To further investigate the influence of illumination on the growth of CNT arrays, we used different carbon sources for comparison. Ethanol was first used for the growth of CNT arrays. After optimization, HACNT arrays with relatively high density can be obtained on the sapphire substrate (see the SEM image in Supplementary Fig. 16a). Raman line mapping measurements were performed on the CNT arrays with illumination, and the results are shown in Supplementary Fig. 16b-c, which indicate there is no obvious selectivity for the semiconducting tubes with illumination. When methane ($CH_4$) was used as the carbon source, the growth of CNTs has a certain window in terms of carbon source supply. The flow rate of methane from 50 to 150 sccm can all be used to obtain HACNT arrays, and amorphous carbon deposition is clearly seen at 150 sccm (Supplementary Fig. 17). The morphology of the arrays grown from FTO catalysts with 50 sccm methane carbon source was further characterized by AFM, demonstrating well aligned HACNT arrays with a density of approximately 10 tubes $\mu m^{-1}$ (Fig. 3e-f). It is worth noting that the density of the array was not affected by the illumination, which excludes the ultraviolet etching effect on CNTs under optimized growth conditions[28]. The Raman line mapping spectra of the arrays under 532 nm laser shown in Fig. 3g indicate the high quality and high purity of CNTs grown on sapphire substrates.

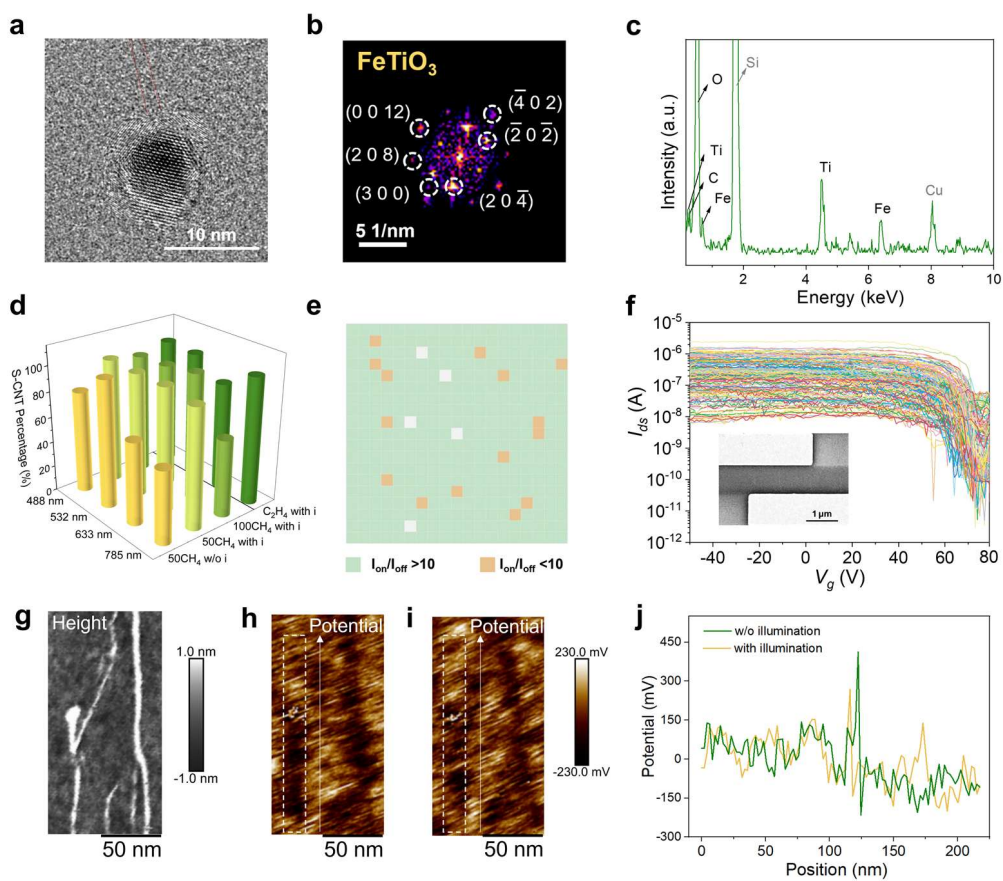

**Fig. 4 Characterization of FTO-samples**. **a-b**, HRTEM image (**a**) and corresponding FFT patterns (**b**) of a catalyst nanoparticle with an CNT grown from it. **c**, EDS results showing the element analysis of the catalysts in mapping region. **d**, A histogram showing the percentage of semiconducting tubes with four different excitation lasers under different growth conditions. **e**, Distribution diagram of FET devices divided by on/off ratio. The white-colour squares stand for invalid devices. **f**, Transfer curves of the bottom-gate FETs showing the semiconducting characters. Inset is a typical SEM image of an FET device. **g-i**, KPFM potential images of a catalyst nanoparticle and grown CNTs without (h) and with (i) illumination, and corresponding height image (g). **j**, The extracted potential signals along the catalyst to the CNT shown in the white box in b-c.

The precise structure of the catalyst was clarified through transmission electron microscopy (TEM). We dispersed a 0.05 mM solution of the FTO precursor on 10 nm $SiO_2$ supports for the standard calcination and growth process, after which the nanoparticles on the supports were analyzed. **Figure 4a** shows a typical high-resolution TEM (HRTEM) image of an individual catalyst nanoparticle with a single-walled CNT grown from it. The corresponding FFT diffraction pattern is demonstrated in Fig. 4b, which is consistent with the FTO crystal in the reciprocal space. This directly proves that FTO is the catalyst for growing carbon nanotubes. A large-area EDS scan was performed, and the results in Supplementary Fig. 18 and Fig. 4c indicate that the catalyst nanoparticles are mainly composed of Fe, Ti, and O, subtracting the signals from supports and systems, which further confirmed the chemical composition of the FTO catalysts. Other catalyst nanoparticles with the same lattice structure of FTO is shown in in Supplementary Fig. 19.

We further performed Raman line mapping tests at four different wavelengths for the CNT arrays under different conditions, in which the spectra of the RBM bands are shown in Supplementary Figs. 20-22. Since there is a clear advantage of semiconducting selectivity under 50 sccm methane conditions, the reference results for this condition without light are placed in Supplementary Fig. 12. For clear comparison, we listed the RBM statistics in Supplementary Table 1, and the corresponding bar chart of semiconducting purity under different excitation lasers is shown in Fig. 4d. Combining the above results, we obtained the following conclusions: Fe catalysts do not have the phenomenon of selective growth under the same illumination conditions, which indicates that the semiconducting-tube enrichment effect of the FTO catalysts is not caused by a simple etching effect, and it is related to the FTO catalysts; when the carbon source is changed to methane, the semiconducting-tube enrichment effect is obvious, indicating that the process is very much related to the carbon source cracking, which initially verified the design of charge transfer; the semiconducting-tube enrichment effect was the best when the supply of the carbon source was small (50 sccm), which means that there existed a rate competition between the normal growth of

CNTs and the selective growth. The above conclusions are of great guidance to our later understanding of the mechanism.

The statistical analysis of electrical properties using Raman spectroscopy has certain limitations because even resonant Raman spectroscopy with three wavelengths cannot cover all chiralities of CNTs. Therefore, to verify the semiconducting CNT enrichment, we transferred the HACNT arrays obtained by growth under the optimal conditions (FTO, 50 sccm methane, under illumination) onto Si/SiO$_2$ (300 nm) substrates by the PMMA-assisted method, and then prepared bottom-gate field effect transistor (FET) device arrays with channel lengths of 1 μm and widths of 2 μm, forming a device array consisting of 20 × 20 = 400 FETs. The electrical properties of the devices were characterized by measuring the transfer curves of the devices using an automatic probe station. As shown in Supplementary Fig. 23 (device array that effectively contact with CNTs contains only 19 × 19 FETs), there are 3-4 tubes in each device channel, as long as one is a metal tube, the on/off current ratio ($I_{on}/I_{off}$) of the device drops below 10. Therefore, we defined devices with on/off ratios greater than 10 as semiconducting-tube devices and those less than 10 as metal-containing devices. With this distinction between the two types of devices, the performance distribution of all valid devices within the array is plotted in Fig. 4e, showing 341 semiconducting devices and 15 metal-containing devices. The transfer curves of the semiconducting-tube devices are demonstrated in Fig. 4f. Using an average density of 3 tubes μm$^{-1}$ as the statistical basis, the test results of the devices indicate that the actual purity of semiconducting carbon nanotubes reaches 98.6%.

**Interpretability of the selective growth**

The above catalyst design and experimental results are based on the ability of the catalyst to inject electrons into CNTs. Therefore, we hope to obtain the explainability of the selective growth of semiconducting tubes through the experimental verification of this phenomenon. Charge transfer is a common phenomenon when materials with different energy band structures come into contact, since both CNTs and catalyst nanoparticles are only a few nanometers in size, we employed Kelvin probe force microscopy (KPFM) to test the potential information of the catalyst and the CNTs to obtain direct proof of charge transfer. We first tried to test the FTO-samples on the sapphire substrate, and the testing tubes were conducted to the substrate by surface plating with gold and silver adhesive as shown in Supplementary Fig. 24. After adjusting the test parameters, the results of the potential distribution were obtained as shown in Supplementary Fig. 25. It is very difficult to quantify the potential difference between the CNTs and catalyst on the substrate due to the strong electrostatic force of the substrate itself, which results in an insignificant potential difference between the CNTs and the catalyst on the substrate. However, since we loaded the driving voltage on the surface of the sample, the surface potential difference ($\Delta\Phi$)= surface potential of the probe ($V_{tip}$) - surface potential of the sample ($V_{Sample}$), and it is obvious in the figure that the surface potential difference at the catalyst is higher than that at the CNTs, which proves that the catalyst's surface potential is lower than that of the carbon nanotubes[29], which is in agreement with our catalyst design. To further eliminate the effect of static electricity on the substrate, we transferred the samples to a conductive silicon wafer for the KPFM test, and the results are shown in Figure 4g-i. Here, we mainly compared the potential difference before and after the illumination at the same location. The extracted potential signals along the catalyst to the CNT (shown in the white box in b-c) are shown in Fig. 4j, indicating that the potential difference between the catalyst and the CNT decreased when illumination is introduced. This data suggests that the light causes a transfer of electrons from the catalyst nanoparticle to the CNTs.

    We can accordingly draw the schematic showing the energy band alignment of the catalyst and the CNT, and the electron transfer. However, it should be noted here that ethanol carbon sources did not achieve selective growth of semiconducting CNTs. As shown in Supplementary Fig. 26a, under neutral conditions, the semiconductor catalyst absorbs light and generates electron-hole pairs, and electrons can be transferred to CNTs from higher energy levels of the catalyst, but the driving force and efficiency of the transfer vary depending on the situation. Obviously, this electron transfer effect is not significant without the built-in electric field (using ethanol as the carbon source). Therefore, the driving force and efficiency of the transfer need to be improved. The

phenomenon of CNT charging during the growth process has been reported previously, which is mainly due to the transfer of electrons from decomposed hydrocarbon-like carbon sources to the catalyst and carbon nanotubes. Therefore, the use of methane or ethylene as the carbon source is equivalent to applying a built-in electric field, which provides the driving force and channel for the subsequent electron transfer. Supplementary Fig. 26b schematically demonstrates the absorption of light, the decomposition of exitons and the electron transfer from the catalyst to the CNT. The holes are hard to transfer because of the energy barrier formed at the contacting interface.

**Conclusions**

In summary, a holistic framework integrating machine learning and traditional methods was constructed, where three key components were utilized: open-access electronic structure databases, the pre-trained natural language processing -based embedding model mat2vec, and physics-driven predictive models for catalyst screening. Based on this framework, a novel method using photocatalysts and light-assisted CVD growth was proposed for the selective growth of semiconducting HACNT arrays. The study evaluated and screened 42 candidate catalysts, and three catalysts with excellent potential were successfully identified. High-throughput automatic experimental verification showed that their semiconducting selectivity exceeded 91%, and the FTO catalyst achieved an impressive 98.6%.

This research not only provides an effective solution for the synthesis of semiconducting CNTs but also, with its AI - assisted global catalyst design approach, holds promise as a general paradigm for addressing complex catalyst design issues in various nanomaterial synthesis scenarios, thus propelling the in-depth development of materials science in terms of precise control and performance optimization.

**Methods**

**Computational details**

All computations were executed based on density functional theory, utilizing the projector augmented-plane wave approach in the Vienna *ab initio* simulation package (VASP). The exchange-correlation functional was described by the Perdew-Burke-Ernzerhof generalized gradient approximation. The treatment of van der Waals forces was handled through the DFT-D3 correction. The plane-wave basis set was defined by a cut-off energy of 400 eV. In the iterative solving process of the Kohn-Sham equations, the convergence threshold for energy was established at $10^{-5}$ eV. A vacuum space of 15 Å was applied normal to the plane to avoid interactions from periodic boundary conditions. Integration over the Brillouin zone was carried out with a 22×1 k-point grid generated by the Monkhorst-Pack method. All geometries were optimized until the magnitudes of the forces on atoms were reduced to under 0.02 eV/Å.

**Preparation of catalyst nanoparticles**

$TiO_2$ and $FeTiO_3$ catalyst precursor solutions were synthesized by the sol–gel method. To get the $FeTiO_3$ precursor, 1 mol $L^{-1}$ Tetra-n-butyl orthotitanate ($Ti(C_4H_9O)_4$) and 1 mol $L^{-1}$ $Fe(NO_3)_2$ $9H_2O$ ethanol solutions were isometrically mixed with addition of 100 μL acetic acid. Yellow and clear mother solution was obtained after stirring for about 12 hours. To get the $TiO_2$ precursor, 0.9 ml ultrapure water was added into tetrabutyl titanate mixed with ethanol solution (1.7 ml in 100 ml ethanol) in a glass reactor under stirring and kept reaction at 70 °C for 4 hours. Fe and $CeO_x$ catalyst precursor solutions were obtained by dissolving $FeCl_3$ and $Ce(NO_3)_3·6H_2O$ in ethanol, respectively. Solutions with concentration of 0.05 mmol $L^{-1}$ were obtained by ethanol dilution, making it suitable as a precursor for nanoparticle formation. After spin-coating (2000 rpm for 60 s) the catalyst precursor solutions onto the substrates, a calcination process under 1100 °C for 8 h was conducted in the air atmosphere, obtaining the oxides catalysts on the substrates, ready for CNT growth or characterization.

**Growth of HACNT arrays**

A-plane sapphire substrates (single-side polished, miscut angle < 0.5°, surface roughness < 0.5 nm) were used for the growth of HACNT arrays, which were acquired from Resea-Tech (Beijing) Co., Ltd. The sapphire substrates underwent annealing at 1100 °C in air for 8h after cleaning, followed by the catalyst loading process mentioned before. The growth process was fully conducted by the home-made automated CVD system (CARCO, carbon copilot). Substrates were conveyed into the mini-CVD (Units Technology Co., Ltd.) by the robotic arm and automated sample holder. Upon the initiation of the growth cycle, the system was programmed to reach the predetermined temperature in 15 minutes. Once the target temperature was achieved, the system sequentially executed the reduction and growth phases. Following the growth cycle, the temperature was rapidly reduced by water-cooling to 150°C. At this point, the samples were automatically retrieved by the sample holder and robotic arm, and the system proceeded to the next set of growth tasks.

The parameter space for the synthesis was defined with specific ranges to optimize the conditions for HACNT array growth. The growth parameters were set as follows: Growth temperature: ranging from 800°C to 1000°C; Reduction time: ranging from 1 to 15 mins; Growth time: 15 mins; Argon flow: 300 sccm; Hydrogen flow: ranging from 200 to 500 sccm; Carbon Source: Ethanol (bubbling by argon flow) ranging from 50 to 100 sccm; methane ranging from 50 to 150 sccm; ethylene ranging from 2 to 10 sccm.

The autonomous optimization was guided by a Bayesian strategy. To minimize the influence of experimental randomness, each iteration was performed in three parallel experiments under identical parameter settings. Samples without CNT growth were excluded, and the average score of the remaining samples was used as feedback for the next iteration. The scoring metric was defined as the product of CNT density and alignment, consistent with our previous study[24].

**Characterization of catalysts**

XPS were performed on an AXIS Supra X-ray Photoelectron Spectrometer (Kratos Analytical Ltd.) with Al K$\alpha$ (1486.6 eV) as the X-ray source. As for the TEM characterization, the preparation of the samples is the same process as growth except for replacing the substrates into 10 nm-thick $SiO_2$ grids. The HRTEM images, ADF-

STEM images as well as EDS mapping were conducted on an aberration-corrected TITAN Cubed Themis G2 300 equipped with Bruker QUANTAX EDS spectrometer under an accelerating voltage of 300 kV. The convergence semi-angle was 21.4 mrad, and collection angle was 39-200 mrad. The beam current was typically about 30 pA when acquiring ADF-STEM images, and a larger beam current value (about 200 pA) was used for EDS mapping to improve the signal-to-noise ratio.

**Characterization of HACNT arrays**

SEM images were obtained on a Hitachi S4800 SEM operated at 1.0 kV. Raman spectra of CNTs were collected by Jovin Yvon-Horiba LabRam systems with four excitation lasers: 488 nm, 532 nm, 633 nm and 785 nm. Line mapping measurements were conducted with a scan step of 1 μm. AFM and KPFM images were obtained using a Dimension Icon microscope (Bruker).

**Devices fabrication and electrical characterization**

The FET device is fabricated as follows. First, the HACNT array is transferred to a 2 cm×2 cm $SiO_2$/Si chip, where the thickness of the $SiO_2$ is 285nm. The CNT transistors are then fabricated by fabricating a periodic array of metallic pads (3/50 nm of Cr/Pd) to contact the CNTs. The channel length between the two metal electrodes is 1μm. As the width of the contact channel is 2 μm, most of the as-fabricated transistors contain more than one CNT.

The electrical measurements were performed at room temperature using a homemade automatic probe station. A data acquisition board (ADwin-Gold II, Jäger Computergesteuerte Messtechnik) is used to apply the bias and gate voltages and read the voltage output of the current–voltage converter (DDPCA-300, Femto Messtechnik). All the devices were measured in a two-terminal setup, where we applied a bias voltage and measured the current.

**Data Availability**

The data that support the findings of this study are available from the corresponding authors upon request. Source data are provided with this paper.

**Code Availability**

The codes that support the findings of this study are available from the corresponding authors upon request.

**Acknowledgements**

L.Q. acknowledges funding from the National Key R&D Program of China (2022YFA1203304) and the National Natural Science Foundation of China (52102032). Jin Z. acknowledges the financial support from the National Key R&D Program of China (2022YFA1203302), the National Natural Science Foundation of China (52021006, 22494641), the Strategic Priority Research Program of CAS (XDB36030100), the Beijing National Laboratory for Molecular Sciences (BNLMS-CXTD-202001), and the Shenzhen Science and Technology Innovation Commission (KQTD20221101115627004). Jian Z. acknowledges funding from the National Natural Science Foundation of China under grant no. 12574047. The authors thank Dr. Guangjie Zhang from National Center for Nanoscience and Technology for his support in KPFM technique.

**Author Contributions Statement**

L.Q. and Y.L. conceived the AI-assisted workflow. Y.L performed the screening of catalysts through ML methods. L.Q. and Y. X. prepared the samples and performed the experiments of growing and characterizing HACNT arrays. Jian Z. performed the fabrication and measurement of CNT-FETs. Y.Y. and Z.L. conducted the STEM-HAADF characterization and the corresponding data analysis. P.L., supervised by F.D., performed the DFT calculations. L.Q. and Y.L. wrote and revised the manuscript with input from all authors. Jin Z. and L.Q. supervised the overall projects. All authors contributed to the discussion and writing of this manuscript.

**Competing Interests Statement**

# Supplementary Information

**Artificial Intelligence-Enabled Holistic Design of Catalysts Tailored for Semiconducting Carbon Nanotube Growth**

Liu Qian*, Yue Li, Ying Xie, Jian Zhang, Pai Li, Yue Yu, Zhe Liu, Feng Ding, Jin Zhang*

**Contents:**

**Suppl. Notes 1-2**

**Supplementary Figs. 1-26**

**Supplementary Tables 1**

**Supplementary References**

**Supplementary Notes**

**Supplementary Note 1 Model-Driven Feature Selection**

The Model-Driven Feature Selection (MDFS) approach systematically evaluates the contribution of various physicochemical descriptors to predict the catalytic behavior during carbon nanotube (CNT) synthesis. By iteratively removing combinations of features and assessing the model performance, this method identifies the most relevant descriptors that capture the underlying mechanisms of CNT growth. The procedure is divided into the following steps, as depicted in Supplementary Fig. 2.

**1.1 Dataset Preparation**

The dataset used in this study comprises approximately 700 high-quality experimental entries, including descriptors of catalyst properties and CNT growth conditions. Growth conditions encompass parameters such as temperature, gas composition, and reaction durations. Catalyst properties include embedding-based descriptors, such as "v_cnt" and "v_catalysis", derived from domain-specific embeddings.

Specifically, to enhance model robustness and interpretability, continuous variables such as temperature, gas flow rates, and reaction times were discretized into predefined intervals, thereby reducing noise. Numerical features were normalized using a RobustScaler, ensuring resistance to the influence of outliers.

As for embedding-based descriptors, we employed a pre-trained word2vec model from the mat2vec package. This model maps chemical formulas and scientific terms into a dense vector space where semantic proximity reflects physicochemical correlations. The embedding-based descriptors were generated by calculating the cosine similarity between the vector representation of each catalyst (e.g., "Fe", "TiO2") and the vectors of specific target keywords. This methodology was uniformly applied to derive two distinct categories of features:

a) Growth-related descriptors: A set of 12 keywords representing critical stages and properties of CNT synthesis, including "nucleation", "cnt", "activation_energy", and so on.

b) Photocatalysis-related descriptor: A specific feature derived from the keyword "photocatalysis" to quantify the material's photocatalytic ability. Notably, while generated using the identical vector-based approach, this descriptor was reserved for quantifying photocatalytic performance and was not included in the feature selection for CNT growth.

**1.2 MLP Model Architecture**

A Multi-Layer Perceptron (MLP) was employed to model the relationship between experimental parameters and CNT growth outcomes. The MLP consists of two fully connected layers with 100 neurons in the hidden layer and 1 neuron in the output layer. ReLU activation functions were applied to introduce non-linearity, and the model was trained to minimize the Mean Squared Error (MSE) between predicted and actual outcomes.

**1.3 Model Training and Evaluation**

The dataset was split into training and testing sets, with 20% allocated for testing (test_size = 0.2). The model was trained for up to 5000 epochs using the Adam optimizer with weight decay to prevent overfitting. An early stopping criterion (patience = 500, min_delta = 0.1) was employed to halt training if no significant improvement was observed in the test loss. Model performance was evaluated using the MSE and R² metrics on both the training and testing datasets, providing a quantitative measure of predictive accuracy and generalizability. The trained model demonstrated robust performance in mapping input experimental conditions and catalyst descriptors to CNT growth efficacy. For each iteration, the following metrics were calculated: Training MSE, Testing MSE, Training $R^2$, Testing $R^2$.

**1.4 Feature Selection Procedure**

The feature selection process aimed to identify the most significant descriptors for modeling CNT growth outcomes. A combinatorial approach was employed to systematically remove features from the dataset and evaluate the impact on model performance. Specifically:

a) All possible feature subsets were generated from a pool of 12 embedding-based descriptors, such as "v_catalysis" and "v_growth".

b) For each subset, the MLP model was retrained, and the predictive performance was evaluated using the testing $R^2$ score.

c) The effectiveness of a given subset was quantified based on improvements or reductions in the $R^2$ metric.

d) Thousands of feature combinations were assessed, and the subsets that yielded the highest testing $R^2$ scores were identified as the most relevant for CNT growth predictions.

This iterative and exhaustive evaluation enabled the identification of descriptors that effectively captured the underlying physicochemical interactions critical to CNT synthesis, guiding the subsequent stages of catalyst screening and optimization.

**Supplementary Note 2: Evaluation of CNT growth ability for various catalysts**

To evaluate the growth ability of candidate catalysts, we first obtained their selected embedding values through the MDFS process using the mat2vec model. Next, a large set of experimental parameters was randomly generated, ensuring coverage across the parameter space within realistic experimental ranges. Using the best-performing model from the MDFS process, we conducted virtual experiments to predict the CNT growth outcomes for each candidate. Finally, the predicted results were averaged to provide a growth ability score, representing the feasibility of achieving high-quality CNT arrays for each catalyst. (Supplementary Fig. 3)

This heatmap in Supplementary Fig. 4 illustrates the correlation matrix of selected descriptors related to the growth of horizontally aligned carbon nanotube (HACNT) arrays and photocatalysis ability. These descriptors were derived by computing the cosine similarity between different catalysts and selected keywords using the mat2vec model. Prior to integrating these descriptors into our predictive model, we conducted a feature importance analysis to assess their relationships and potential redundancies.

Each cell represents the Pearson correlation coefficient between two descriptors, ranging from -1 to 1, where positive values indicate a direct relationship and negative values suggest an inverse correlation. The color gradient reflects the correlation strength: warmer colors (red) represent strong positive correlations, whereas cooler colors (blue) denote negative correlations.

From the heatmap, we observe that while some descriptors show moderate correlation (e.g., v_growth and v_nucleation), most features remain relatively independent. This suggests that the selected descriptors provide diverse and complementary information about catalyst behavior. Importantly, the lack of extreme correlations ensures that these features contribute unique insights to the model rather than introducing redundancy. This step validates the effectiveness of our descriptor selection approach before their application in catalyst screening and CNT growth predictions.

The embedding-derived descriptors, obtained through cosine similarity calculations using the mat2vec model, capture high-level physicochemical relationships informed by literature knowledge. In contrast, the experimental parameters, such as temperature, gas composition, and reaction time, are direct process variables that govern CNT synthesis.

The heatmap in Supplementary Fig. 5 demonstrates that these two groups of features occupy separate feature spaces, with minimal correlation between them. This independence suggests that the embedding-derived descriptors provide unique information that is not inherently captured by traditional experimental parameters. This distinction is crucial for model performance, as it ensures that the inclusion of embedding-based features enriches the predictive framework without introducing redundant information. Furthermore, the observed independence underscores the potential of leveraging AI-driven feature extraction techniques to complement conventional material screening methodologies, enabling a more comprehensive evaluation of catalyst suitability for CNT growth.

Supplementary Fig. 6a presents a scatter plot comparing the predicted and actual growth scores for both the training (red) and test (blue) datasets. The model demonstrates strong predictive capabilities, as evidenced by the close alignment of data points with the diagonal dashed line, which represents an ideal prediction. The inset plot displays the Mean Squared Error (MSE) loss as a function of training epochs, showing that the model reaches a stable state after approximately 200 epochs without indications of overfitting.

Supplementary Fig. 6b provides insights into feature importance using SHAP (Shapley Additive Explanations) values, which quantify each feature's impact on the model output. Features such as growth conditions (v_growth, temperature, $H_2$ concentration, and v_carbon) exhibit significant influence, highlighting their direct correlation with CNT synthesis outcomes. The color gradient represents the relative feature values, with blue indicating lower values and red indicating higher values. Notably, v_growth and temperature have the strongest positive influence on CNT formation, while some gas-phase parameters and catalyst descriptors contribute to a lesser extent. This analysis confirms the model's ability to extract meaningful relationships between experimental conditions, catalyst descriptors, and CNT growth performance.

**Supplementary Figures**

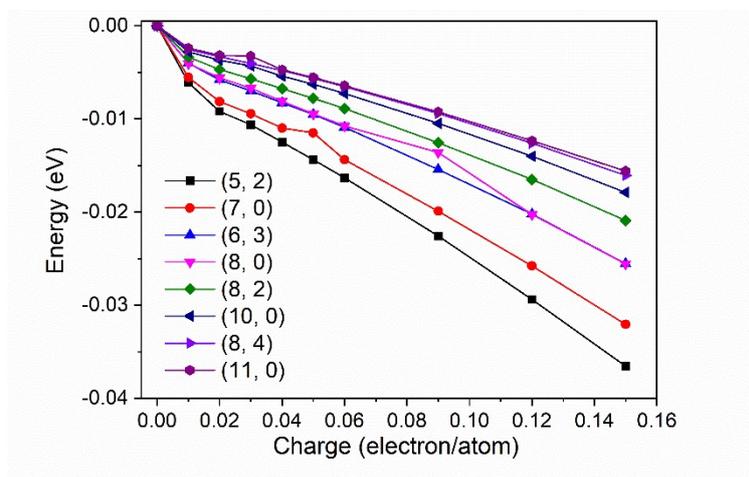

**Supplementary Fig. 1** Diagram of the CNT energy change with the accumulation of electrons.

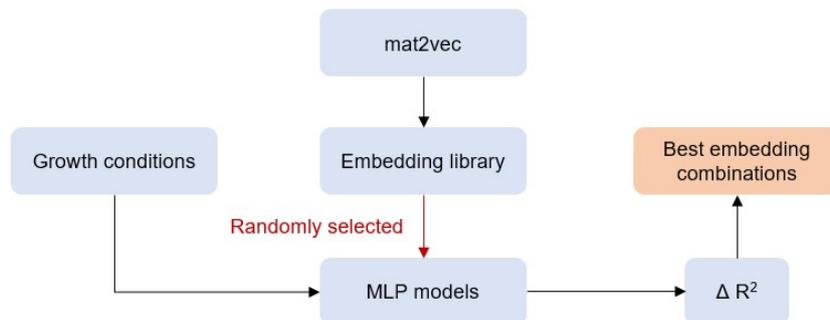

**Supplementary Fig. 2** The architecture for Model-Driven Feature Selection.

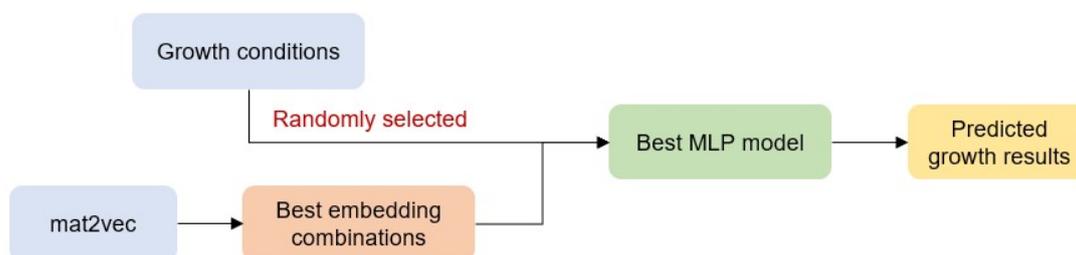

**Supplementary Fig. 3** The architecture for evaluation of CNT growth ability for various catalysts.

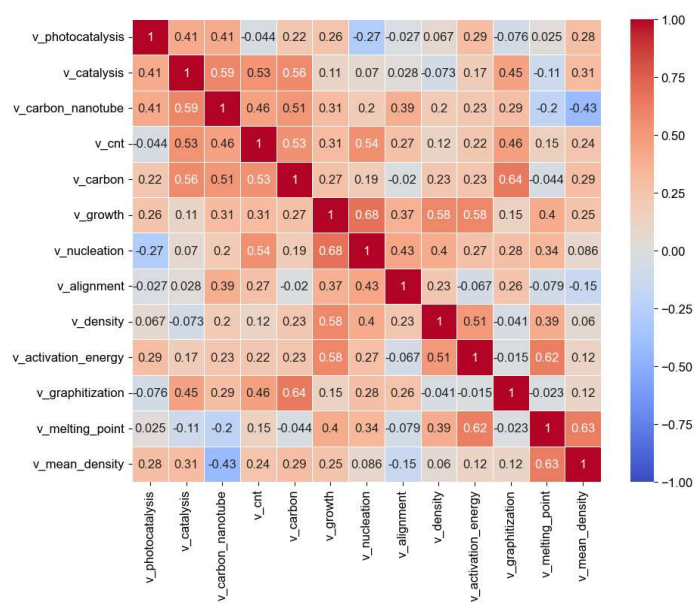

**Supplementary Fig. 4** Correlation heatmap of the 13 selected descriptors (including v_photocatalysis and synthesis-related descriptors) derived from mat2vec embeddings.

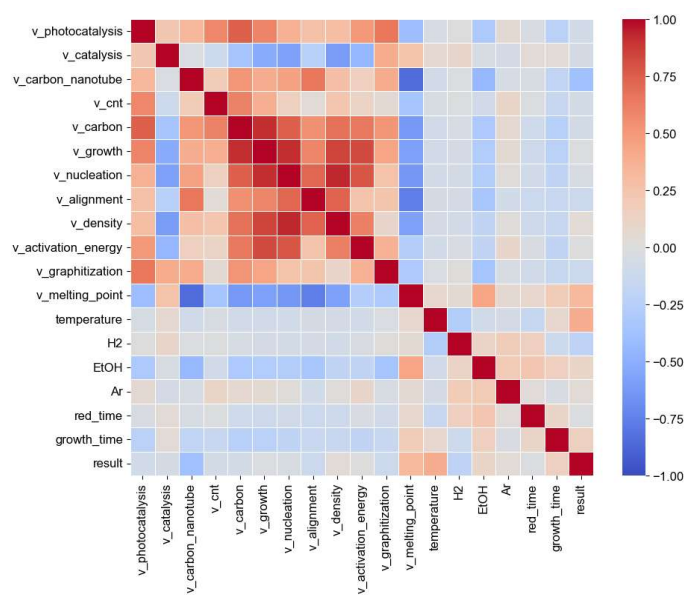

**Supplementary Fig. 5** Correlation heatmap of embedding-derived descriptors and experimental growth parameters.

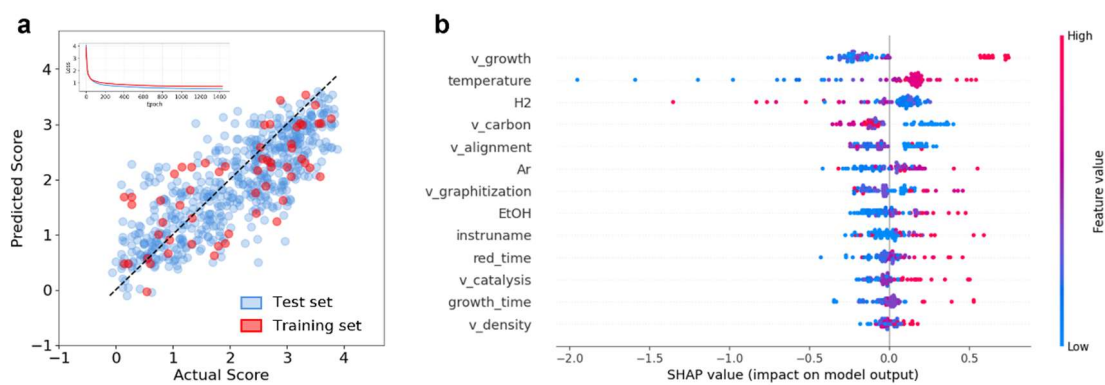

**Supplementary Fig. 6 a**, Model performance evaluation through predicted vs. actual scores. The inset shows the training loss (MSE) over epochs. **b**, SHAP analysis of feature contributions to the model output.

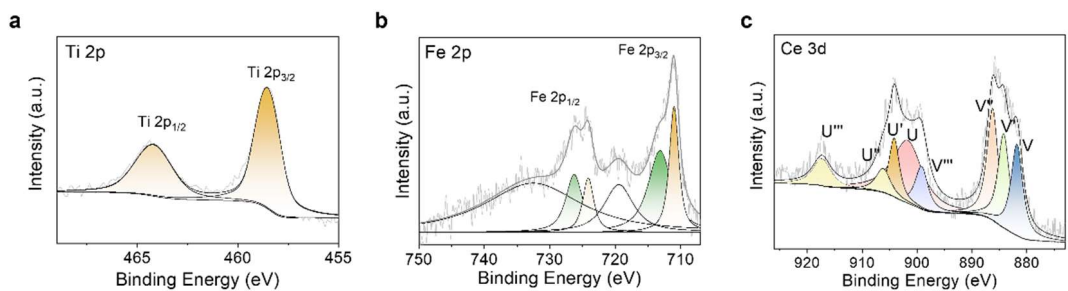

**Supplementary Fig. 7 a**, XPS spectra of a TiO$_2$-sample showing the Ti 2p region with two peaks of Ti$^{4+}$. **b**, XPS spectra of an FTO-sample showing the Fe 2p region with Fe$^{2+}$ and Fe$^{3+}$ peaks. **c**, XPS spectra of a CeO$_x$-sample showing the Ce 3d region with multiple peaks of several Ce oxidation states.

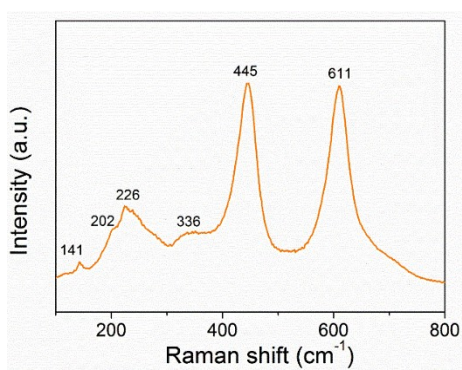

**Supplementary Fig. 8** Raman spectrum of the FTO-sample after calcination.

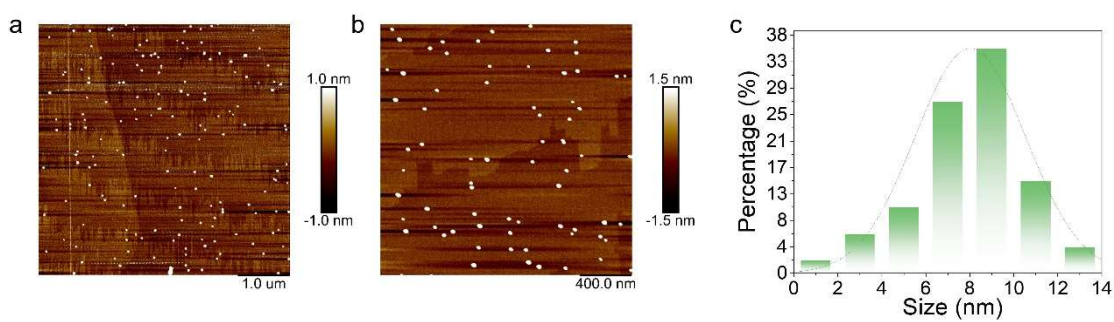

**Supplementary Fig. 9 a-b**, AFM images of catalyst nanoparticles. **c**, Size statistics of catalyst nanoparticles.

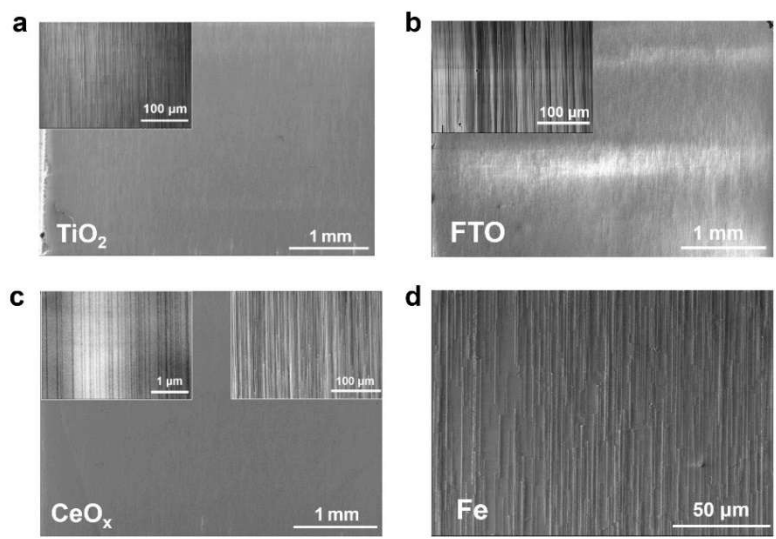

**Supplementary Fig. 10** SEM images with different magnifications of HACNT arrays grown from different catalysts.

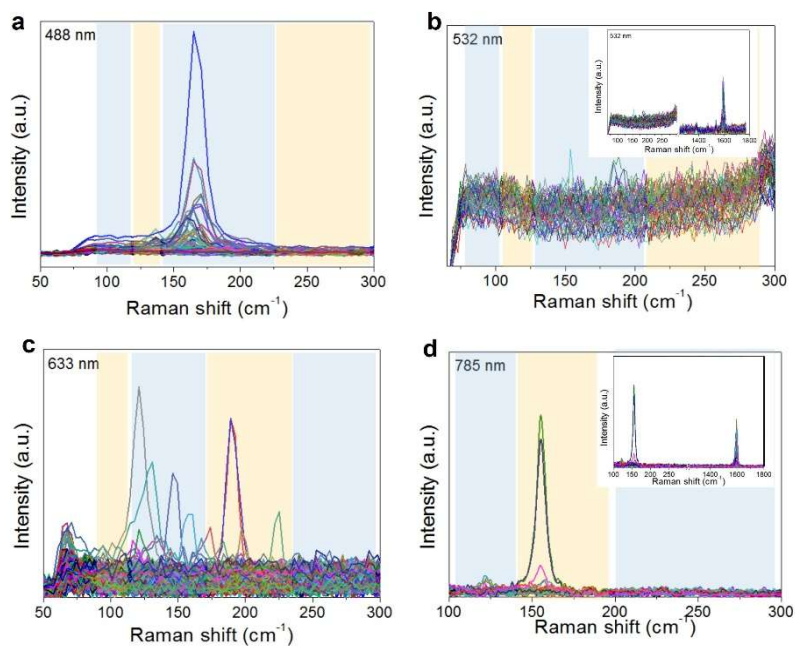

**Supplementary Fig. 11** Raman spectra of HACNT arrays grown from FTO catalysts without illumination under different excitation wavelengths (50 sccm $CH_4$). **a**, 488 nm. **b**, 532 nm. **c**, 633 nm. **d**, 785 nm.

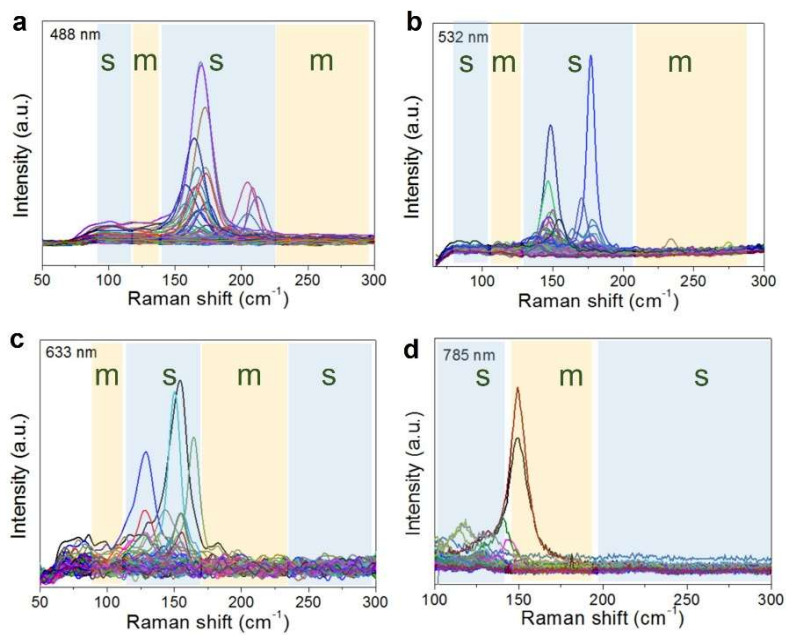

**Supplementary Fig. 12** Raman spectra of HACNT arrays grown from FTO catalysts with illumination under different excitation wavelengths (50 sccm CH$_4$). **a**, 488 nm. **b**, 532 nm. **c**, 633 nm. **d**, 785 nm. The blue and yellow background represents peaks of semiconducting tubes and metallic tubes respectively.

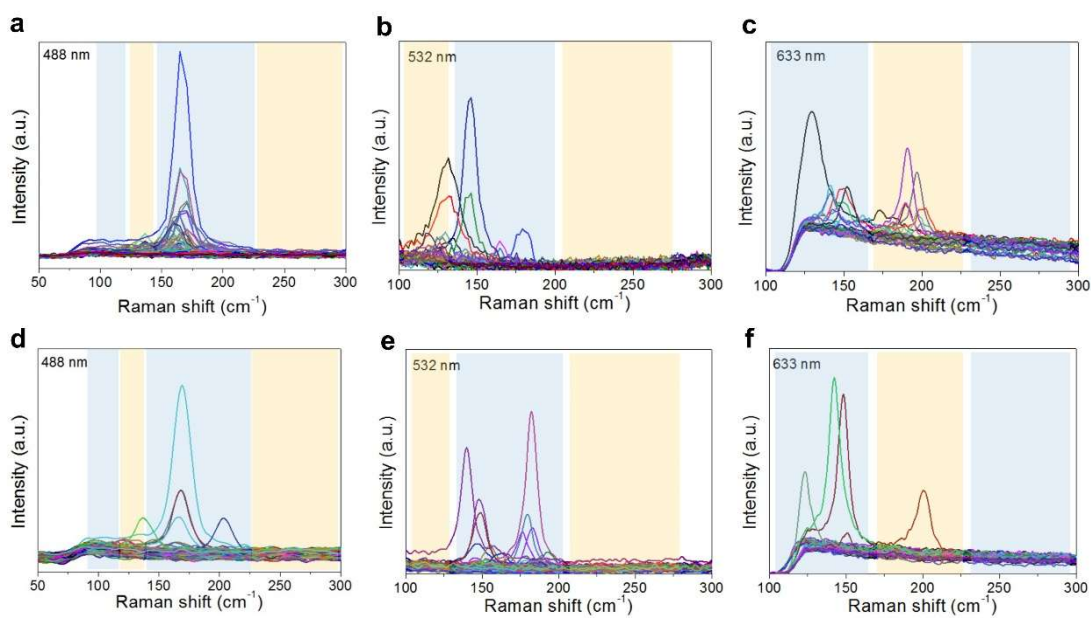

**Supplementary Fig. 13** Raman spectra of HACNT arrays grown from TiO$_2$ catalysts without (**a-c**) and with (**d-f**) illumination under different excitation wavelengths (50 sccm CH$_4$). **a,d,** 488 nm. **b,e**, 532 nm. **c,f,** 633 nm.

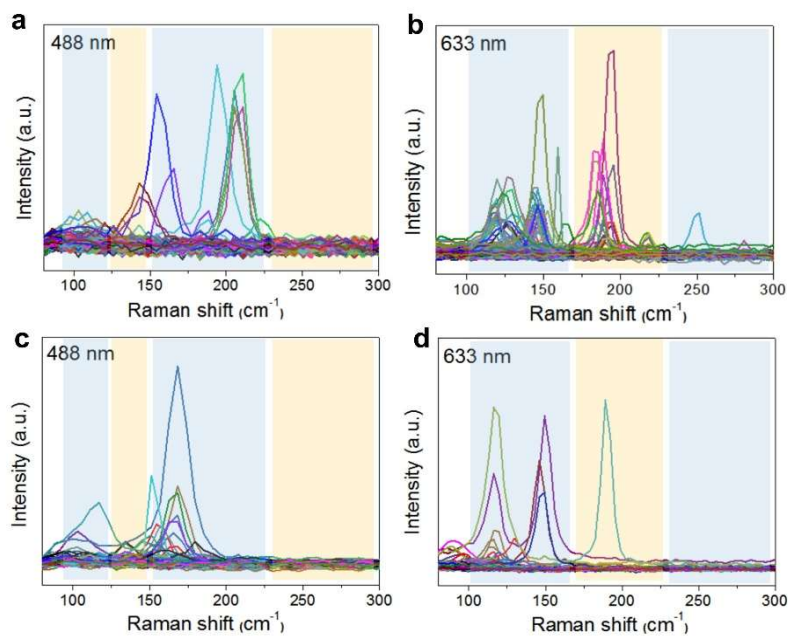

**Supplementary Fig. 14** Raman spectra of HACNT arrays grown from CeO$_x$ catalysts without (**a-b**) and with (**c-d**) illumination under different excitation wavelengths (65 sccm CH$_4$). **a,c,** 488 nm. **b,d**, 633 nm.

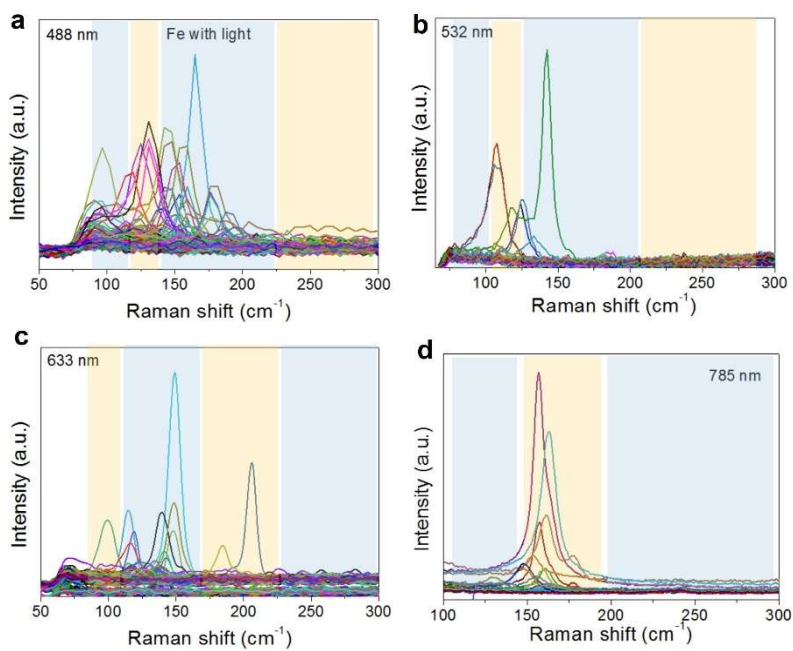

**Supplementary Fig. 15** Raman spectra of HACNT arrays grown from Fe catalysts with illumination under different excitation wavelengths (50 sccm CH$_4$). **a**, 488 nm. **b**, 532 nm. **c**, 633 nm. **d**, 785 nm.

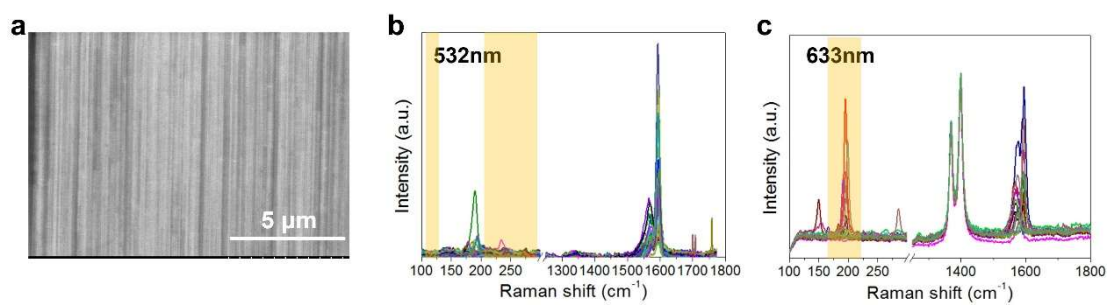

**Supplementary Fig. 16 Characterization of CNT arrays grown from FTO catalysts with ethanol as carbon source. a,** A typical SEM image of the CNT arrays. **b-c**, Raman line mapping spectra of the arrays excited by 532 nm laser (b) and 633 nm laser (c). The yellow bars represent the RBM range of metallic tubes.

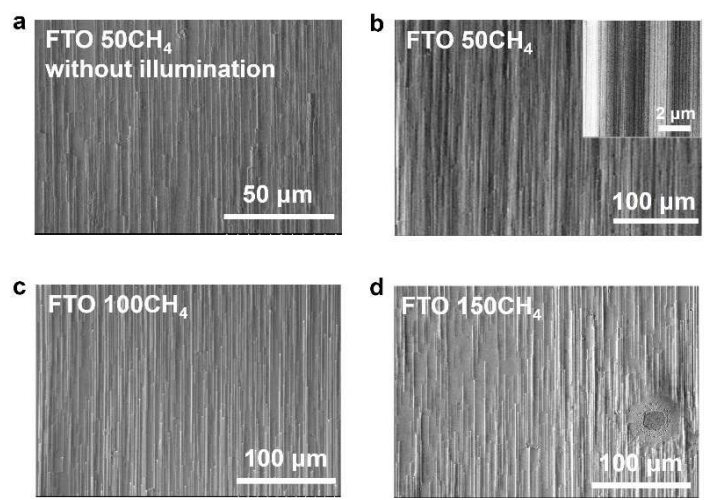

**Supplementary Fig. 17** SEM images of HACNT arrays grown from FTO catalysts with $CH_4$ as the carbon source under different growth conditions. **a**, 50 sccm $CH_4$ without illumination. **b**, 50 sccm $CH_4$ with illumination. **c**, 100 sccm $CH_4$ with illumination. **d**, 150 sccm $CH_4$ with illumination.

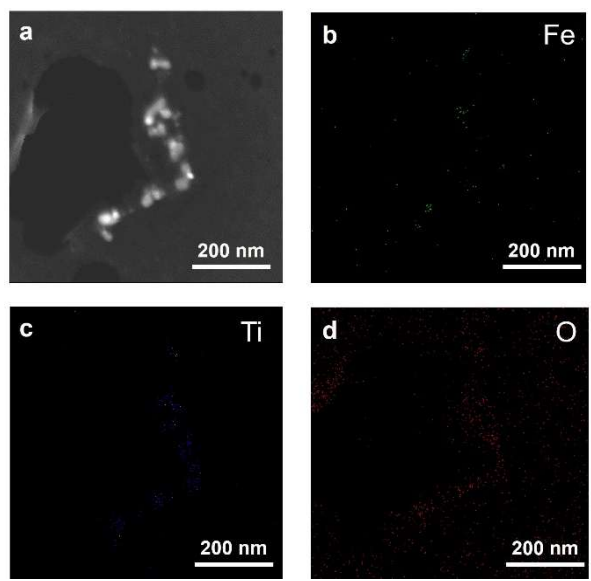

**Supplementary Fig. 18** EDS element scan images of a group of catalyst nanoparticles. **a**, The STEM image of the catalyst nanoparticles. **b**, Corresponding Fe-element scan image with green colour. **c**, Corresponding Ti-element scan image with blue colour. **d**, Corresponding O-element scan image with red colour.

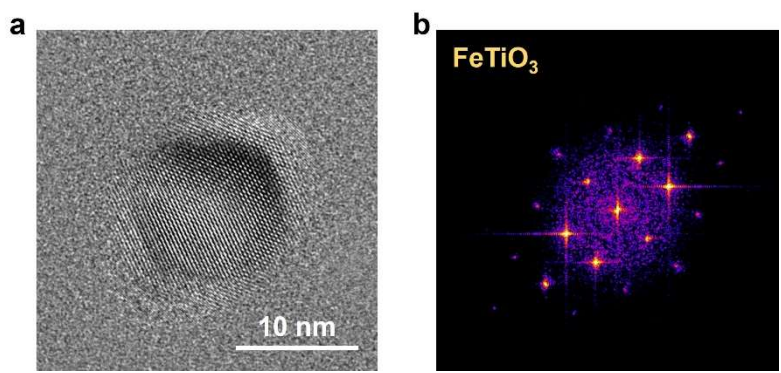

**Supplementary Fig. 19** HRTEM image (**a**) and corresponding FFT patterns (**b**) of a catalyst nanoparticle.

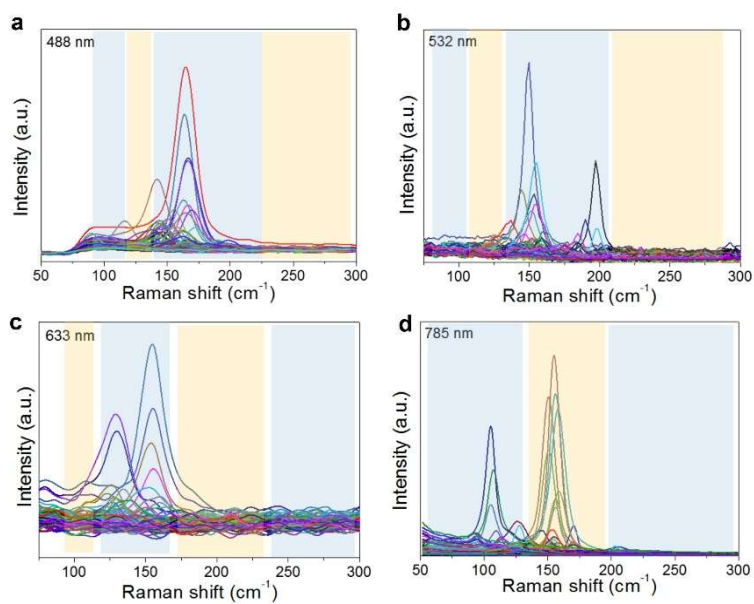

**Supplementary Fig. 20** Raman spectra of HACNT arrays grown from FTO catalysts with illumination under different excitation wavelengths (100 sccm CH$_4$). **a**, 488 nm. **b**, 532 nm. **c**, 633 nm. **d**, 785 nm.

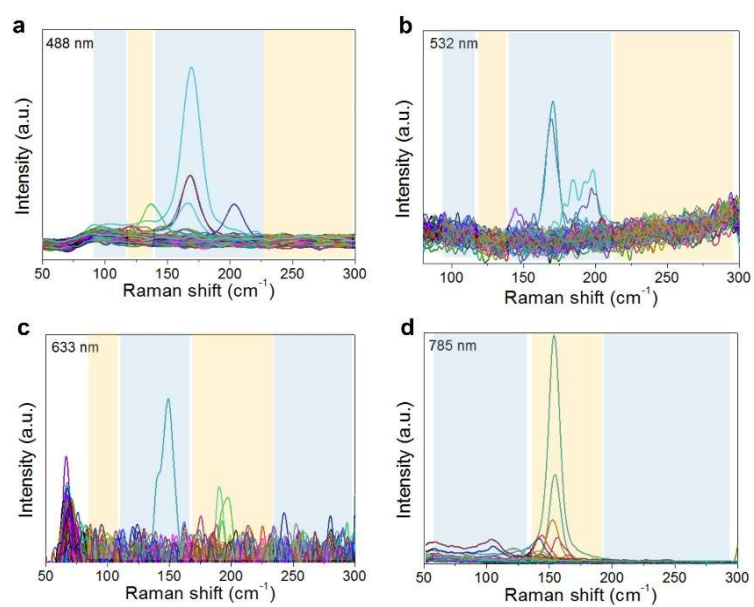

**Supplementary Fig. 21** Raman spectra of HACNT arrays grown from FTO catalysts with illumination under different excitation wavelengths (150 sccm CH$_4$). **a**, 488 nm. **b**, 532 nm. **c**, 633 nm. **d**, 785 nm.

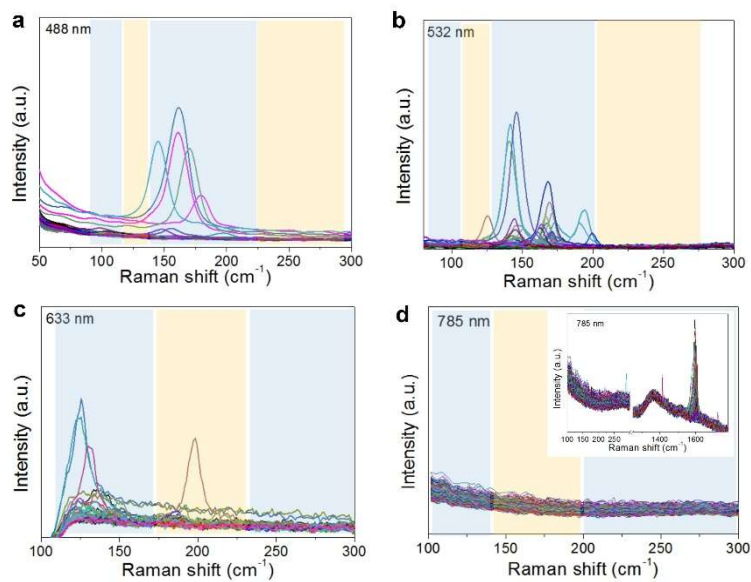

**Supplementary Fig. 22** Raman spectra of HACNT arrays grown from FTO catalysts with illumination under different excitation wavelengths (4 sccm $C_2H_4$). **a**, 488 nm. **b**, 532 nm. **c**, 633 nm. **d**, 785 nm.

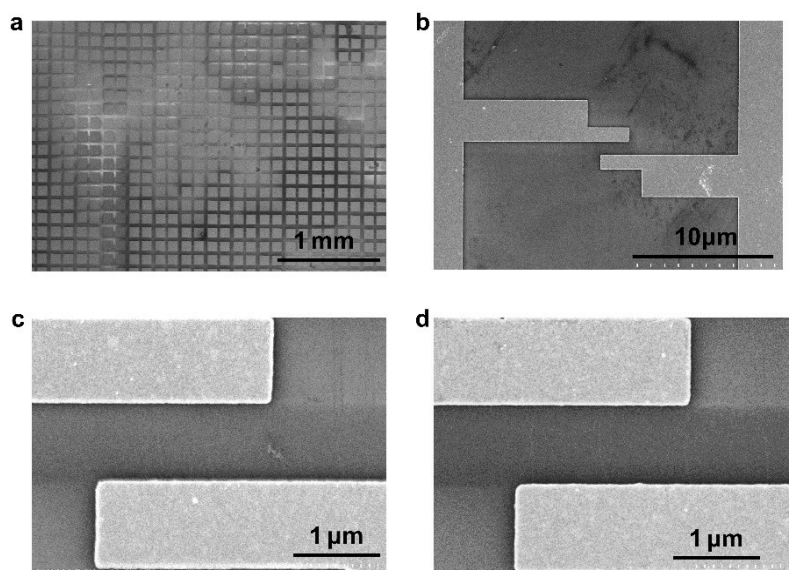

**Supplementary Fig. 23** SEM images with different magnifications of FET devices based on our HACNT arrays.

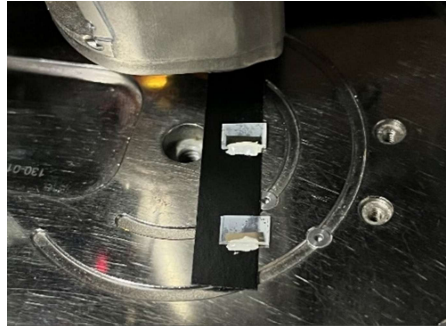

Supplementary Fig. 24 Picture of KPFM test on sapphire substrate.

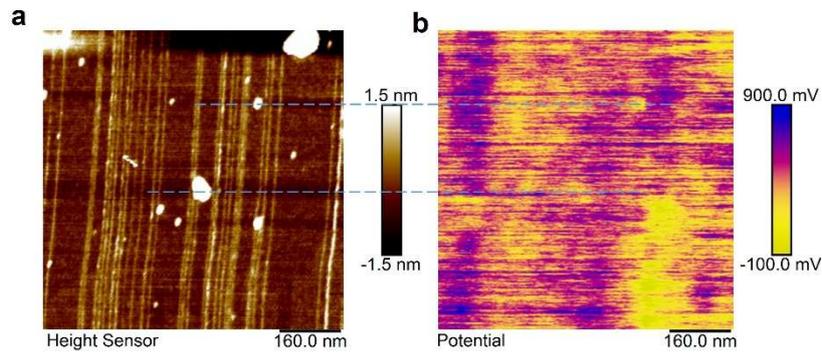

Supplementary Fig. 25 KPFM height image (a) and corresponding potential image (b) of carbon nanotubes and catalyst nanoparticles on sapphire substrate.

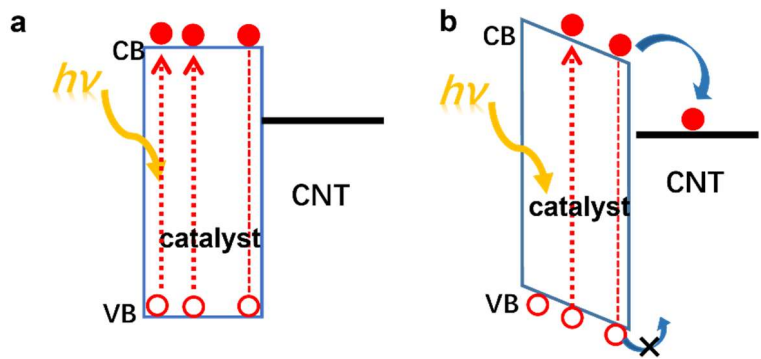

Supplementary Fig. 26 Schematics showing the situations without (a) and with (b) built-in electric field.

**Supplementary Table 1** RBM peak statistics of Raman line mapping tests for CNTs grown under different conditions

| Number of RBMs | 488 nm | | 532nm | | 633nm | | 785 nm | | Content of s-CNTs |
|---|---|---|---|---|---|---|---|---|---|
| | s | m | s | m | s | m | s | M | |
| FTO 50CH$_4$ w/o i | 77 | 19 | 5 | 0 | 24 | 13 | 8 | 6 | 75% |
| FTO 50CH$_4$ with i | 116 | 2 | 152 | 4 | 39 | 1 | 60 | 4 | 97% |
| FTO 100CH$_4$ with i | 52 | 3 | 75 | 3 | 21 | 0 | 20 | 13 | 91% |
| FTO C$_2$H$_4$ with i | 13 | 0 | 83 | 1 | 11 | 2 | 0 | 0 | 97% |
| Fe 50CH$_4$ with i | 64 | 18 | 5 | 5 | 41 | 10 | 4 | 23 | 67% |